\documentclass[twocolumn,epsfig,showpacs,floatfix,superscriptaddress,preprintnumbers,amsmath,amssymb]{revtex4}
\usepackage{amssymb}
\usepackage{amsmath}
\usepackage{latexsym}
\usepackage{graphicx}
\usepackage{dcolumn}
\usepackage{bm}
\DeclareGraphicsExtensions{.eps}

\newcommand{\be}{\begin{equation}}
\newcommand{\ee}{\end{equation}}
\newcommand{\ba}{\begin{eqnarray}}
\newcommand{\ea}{\end{eqnarray}}
\newcommand{\baa}{\begin{eqnarray*}}
\newcommand{\eaa}{\end{eqnarray*}}

\begin{document}

\title{Generalized Schrieffer-Wolff Transformation of 2 Kondo Impurity Problem}
\author{T. Tzen Ong}
\affiliation{Department of Applied Physics, Stanford University, Stanford CA 94305, USA}
\affiliation{IBM Research Division, Almaden Research Center, San Jose, CA 95120, USA}
\author{B. A. Jones}
\affiliation{IBM Research Division, Almaden Research Center, San Jose, CA 95120, USA}
\date{\today}

\begin{abstract}
We have carried out a generalized Schrieffer-Wolff transformation of an Anderson two-impurity Hamiltonian to study the low-energy spin interactions of the system. The $2^{nd}$-order expansion yields the standard Kondo Hamiltonian for two impurities with additional scattering terms. At $4^{th}$-order, we get the well-known RKKY interaction. In addition, we also find an antiferromagnetic superexchange coupling and a correlated Kondo coupling between the two impurities.
\end{abstract}
\maketitle

\section{Introduction}
\label{sec:intro}
The low-energy interactions of a system of magnetic impurities coupled to a sea of conduction electrons display a rich variety of physical phenomena. One of the most well-known many-body effects is screening of the local moment by the conduction electrons, first explained by Jun Kondo \cite{JKondo}. The Kondo effect was first observed in measurements of dilute magnetic impurities in metals \cite{Hewson}, and is now ubiquitious in many systems including quantum dots (QD) \cite{DGG98Sc}, \cite{DGG98PRl} and magnetic atoms in surfaces probed by scanning tunneling microscopy \cite{Crommie}.

In materials with a dense lattice of magnetic impurities, there is an interesting interplay between the Kondo effect and magnetic ordering of the local moments, which is of particular interest in the heavy-fermion compounds. As the simplest manifestation of this competition, and as a first step towards understanding the lattice problem, considerable attention has been placed on the two Kondo impurity problem. The problem was first solved exactly in the particle-hole symmetric case using NRG \cite{Jones87} and subsequently by the Bethe ansatz, conformal field theory \cite{Affleck}, and bosonization \cite{Gan}. A second-order phase transition is found for the case of particle-hole symmetry, which typically turns into a cross-over when potential scattering is introduced.

\begin{figure}[bht]
\begin{center}
\includegraphics[angle=-90,width=8cm]{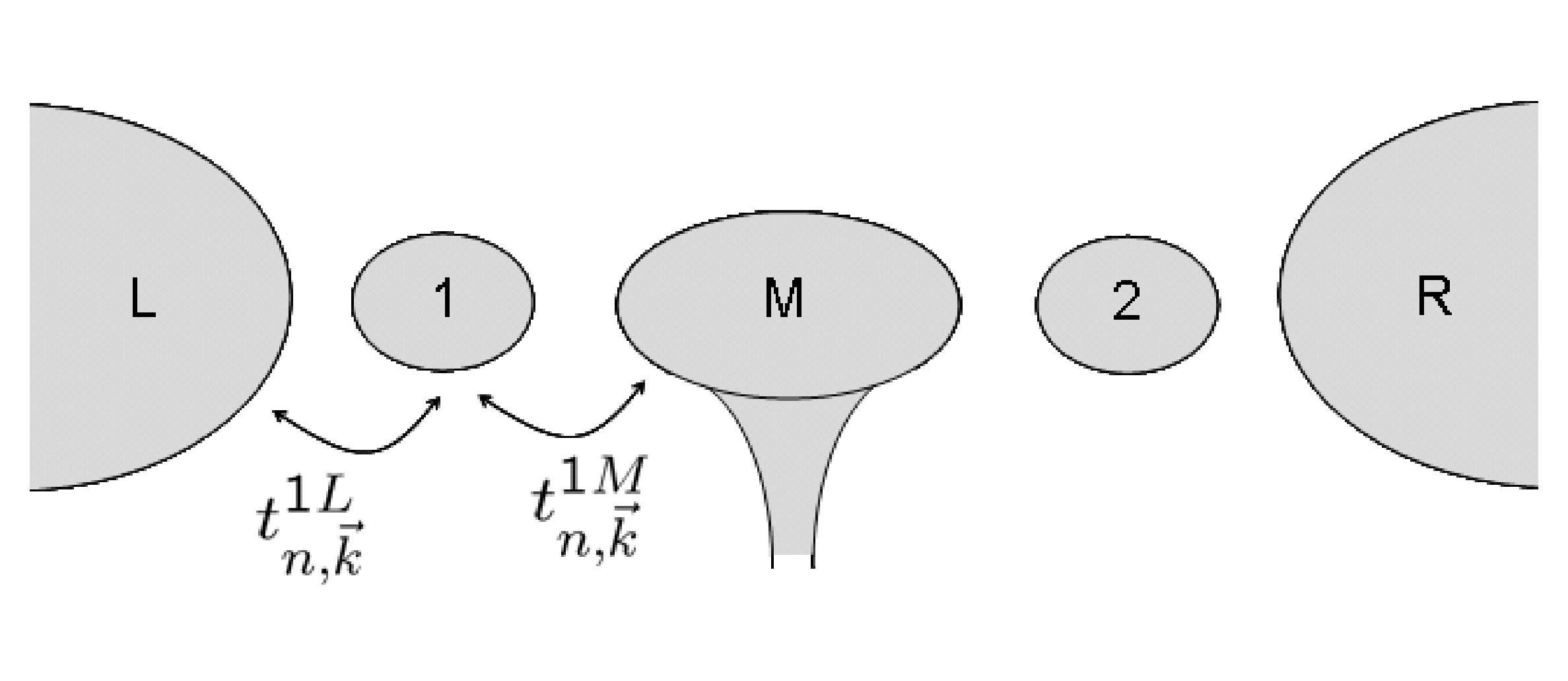}
\end{center}
\caption{The magnetic impurities and the electron reservoirs are labeled 1 and 2, and L, M and R respectively. The tunneling between the localized electron on impurity 1 and the conduction electrons  in the left and middle reservoirs are shown as $t^{1,L}_{n,\vec{k}}$ and $t^{1,M}_{n,\vec{k}}$.}
\label{2Dot_layout}
\end{figure}

A recent two quantum dot experiment by Craig et al. \cite{Marcus} in the geometry shown in Fig. \ref{2Dot_layout} showed a change in the Kondo resonance of one dot by tuning the second dot from an even to an odd occupation. It has been suggested that this is due to RKKY coupling between the dots. We have looked at the effective inter-dot interactions obtained via a canonical transformation of the two impurity Anderson model. The purpose of the calculation is to extract the low-energy spin excitations of the system in order to predict the magnetic interactions. We expect such interactions to occur in STM systems of magnetic impurities on surfaces as well.

\section{Anderson Model of QD}
\label{sec:model}
The interactions of a quantum dot (QD) with its leads can be modelled using the Anderson Hamiltonian \cite{PALee88}, \cite{PALee93}, \cite{Wingreen94}. For a two impurity system in the geometry as shown in Fig. 1, we propose a two impurity Anderson Hamiltonian.

\ba
\mathcal{H} & = & \sum_{\alpha = L,M,R}\sum_{\vec{k},\sigma} \epsilon_{\alpha,\vec{k}} \ c_{\alpha,\vec{k},\sigma}^{\dagger} c_{\alpha,\vec{k},\sigma} \nonumber \\
 & & + \sum_{i=1,2} \sum_{n,\sigma} \epsilon_{i,n} \ d_{i,n,\sigma}^{\dagger} d_{i,n,\sigma} + U_{i} (\hat{n}_i - N_i)^2 \nonumber \\
 & & + \sum_{\alpha=L,M} \sum_{n,\vec{k},\sigma} \left[ t^{1,\alpha}_{n,\vec{k}}  \ e^{-i \vec{k} \cdot \vec{r_1}} \ c_{\alpha,\vec{k},\sigma}^{\dagger} d_{1,n,\sigma} \right. \nonumber \\
 & & \left. \;\;\;\;\;\;\;\;\;\;\;\;\;\;\;\;\; + \, (t^{1,\alpha}_{n,\vec{k}})^* \ e^{i \vec{k} \cdot \vec{r_1}} \ d_{1,n,\sigma}^{\dagger} c_{\alpha,\vec{k},\sigma} \right] \nonumber \\
 & & + \sum_{\alpha=M,R} \sum_{n,\vec{k},\sigma} \left[ t^{2,\alpha}_{n,\vec{k}}  \ e^{-i \vec{k} \cdot \vec{r_2}} \ c_{\alpha,\vec{k},\sigma}^{\dagger} d_{2,n,\sigma} \right. \nonumber \\
 & & \left. \;\;\;\;\;\;\;\;\;\;\;\;\;\;\;\;\; + \, (t^{2,\alpha}_{n,\vec{k}})^* \ e^{i \vec{k} \cdot \vec{r_2}} \ d_{2,n,\sigma}^{\dagger} c_{\alpha,\vec{k},\sigma} \right]
\label{Anderson H}
\ea

Here, $\epsilon_{\alpha,\vec{k}}$ is the kinetic energy of the conduction electrons in the leads ($\alpha =$ L, M and R), and $t^{1,\alpha}_{n,\vec{k}}$ and $t^{2,\alpha}_{n,\vec{k}}$ are the tunneling rates between the localized d-electrons at sites 1 and 2, and conduction electrons at lead $\alpha$. $U_i$ and $\epsilon_i$ are the Coulomb charging energies and single-particle energies of the localized electrons in the two QDs.

The low-energy spin interactions of the system are obtained from a generalized form of the Schrieffer-Wolff transformation, carried out on the Anderson two impurity Hamiltonian. The standard Schrieffer-Wolff transformation \cite{Schrieffer-Wolff} of a a single Anderson impurity Hamiltonian maps it to a single Kondo impurity problem using a canonical transform with an operator $\mathbf {S}$. The generalized Schrieffer-Wolff transformation, similar to a canonical transformation for the Hubbard model carried out by MacDonald et al \cite{MacDonald}, is defined in the following manner.
\ba
\tilde{H} & \equiv & \exp^{S} H \exp^{-S} \nonumber \\
S & = & \sum_i S^{(i)}
\label{SW Transform}
\ea
where $S^{(i)}$ is of order $O(\frac{t^i}{U^{i-1}})$. $S_i$ is determined self-consistently at each level of the expansion as shown below. The usual Kondo Hamiltonian is obtained at $2^{nd}$ order in this expansion, and we carry out this expansion to $4^{th}$ order to obtain the Ruderman-Kittel-Kasuya-Yosida (RKKY) and any other interactions at this order. Since we are only interested in the low-energy physics, we will discard terms that take the system out of the ground-state manifold, i.e. terms that cost energy $U$. This gives the following requirement at each level of expansion.
\be
(\mathcal{I} - P_G) \ \tilde{H} \ P_G = 0
\label{PG}
\ee
where $P_G$ is the Gutzwiller projector onto the ground state, which in this case is the singly-occupied d-state on both local sites. At first-order, the following condition is the defining equation for $S^{(1)}$.
\be
(\mathcal{I} - P_G) \ \left( H_I + [S^{(1)},H_0] \right) \ P_G = 0
\label{S1 Def}
\ee
A straight-forward calculation then gives $S^{(1)}$.
\ba
S^{(1)} & = & \sum_{\alpha=L,M} \sum_{\vec{k},\sigma}\left( \frac{1-n_{1,-\sigma}}{\epsilon_{\alpha,\vec{k}} - \epsilon_1} + \frac{n_{1,-\sigma}}{\epsilon_{\alpha,\vec{k}} - \epsilon_1 - U_1}\right) \nonumber \\ 
 & & \times \left( t^{1,\alpha}_{\vec{k}} \ e^{-i \vec{k} \cdot \vec{r_1}} \ c_{\alpha,\vec{k},\sigma}^{\dagger} d_{1,\sigma} \right. \nonumber \\
 & & + \left. (t^{1,\alpha}_{\vec{k}})^* \ e^{i \vec{k} \cdot \vec{r_1}} \ d_{1,\sigma}^{\dagger} c_{\alpha,\vec{k},\sigma} \right)
 \label{S1}
\ea
Here, we have dropped the subscript $n$ for the single-particle level in each QD as we are only interested in the highest level near $E_F$. $S^{(1)}$ is anti-hermitian as required so that $\tilde{H}$ will be hermitian. The effective Hamiltonian at $2^{nd}$-order is given by $\tilde{H} = H_0 + \frac{1}{2}[S^{(1)},H_I]$, where we find the Kondo terms for dot 1 and 2, shown in Eq. \ref{Heff 2}.
\begin{widetext}
\ba
\tilde{H} & = & \sum_{\alpha = L,M,R}\sum_{\vec{k},\sigma} \epsilon_{\alpha,\vec{k}} \ c_{\alpha,\vec{k},\sigma}^{\dagger} c_{\alpha,\vec{k},\sigma} + \sum_{\alpha,\alpha '=L,M} \sum_{\vec{k},\vec{k'}} J1^{\alpha,\alpha '}_{\vec{k},\vec{k'}} e^{i(\vec{k} - \vec{k'}) \cdot \vec{r_1}} \vec{s}^{\alpha,\alpha '}_{\vec{k},\vec{k'}} \cdot \vec{S}_1 + \sum_{\alpha,\alpha '=M,R} \sum_{\vec{k},\vec{k'}} J2^{\alpha,\alpha '}_{\vec{k},\vec{k'}} e^{i(\vec{k} - \vec{k'}) \cdot \vec{r_2}} \vec{s}^{\alpha,\alpha '}_{\vec{k},\vec{k'}} \cdot \vec{S}_2 \nonumber \\
 & & + \sum_{\alpha,\alpha '=L,M} \sum_{\vec{k},\vec{k'}} \frac{1}{2}(K1^{\alpha,\alpha '}_{\vec{k},\vec{k'}} - \frac{1}{2}J1^{\alpha,\alpha '}_{\vec{k},\vec{k'}} n_1) e^{i(\vec{k} - \vec{k'}) \cdot \vec{r_1}} c_{\alpha,\vec{k},\sigma}^{\dagger} c_{\alpha,\vec{k'},\sigma} \nonumber \\
 & & + \sum_{\alpha,\alpha '=M,R} \sum_{\vec{k},\vec{k'}} \frac{1}{2}(K2^{\alpha,\alpha '}_{\vec{k},\vec{k'}} - \frac{1}{2}J2^{\alpha,\alpha '}_{\vec{k},\vec{k'}} n_1) e^{i(\vec{k} - \vec{k'}) \cdot \vec{r_2}} c_{\alpha,\vec{k},\sigma}^{\dagger} c_{\alpha,\vec{k'},\sigma} \nonumber \\
 & & + \sum_{\alpha=L,M} \sum_{\vec{k},\sigma} \frac{1}{2}(J1^{\alpha,\alpha '}_{\vec{k},\vec{k'}} n_{1,-\sigma} - K1^{\alpha,\alpha '}_{\vec{k},\vec{k'}})n_{1,\sigma} + \sum_{\alpha=M,R} \sum_{\vec{k},\sigma} \frac{1}{2}(J2^{\alpha,\alpha '}_{\vec{k},\vec{k'}} n_{2,-\sigma} - K2^{\alpha,\alpha '}_{\vec{k},\vec{k'}})n_{2,\sigma}
\label{Heff 2}
\ea
\end{widetext}
Here, $J1^{\alpha,\alpha '}_{\vec{k},\vec{k'}}$ and $J2^{\alpha,\alpha '}_{\vec{k},\vec{k'}}$ are the Kondo coupling at QD 1 and 2 respectively. The next two terms represent potential scattering at the two sites, and are important at the quantum cross-over point between Kondo-dominated and RKKY-dominated behaviour. The expressions for $J$ and $K$ in terms of the original parameters are given in Appendix \ref{app:Kondo Def}. The last two terms simply renormalize the single particle energies of the local d-electrons.
There are also double occupancy and double vacancy terms that cost energy $U$, but are cancelled at $2^{nd}$-order by $[S^{(2)},H_0]$, as required by the Gutzwiller projection.

At $3^{rd}$-order, all the terms generated by the commutators of $H_{int}$, $S^{(1)}$ and $S^{(2)}$ take the system out of the ground state, and hence are required to cancel completely with $[S^{(3)},H_0]$. However, these terms generate terms at $4^{th}$-order that give rise to the well-known RKKY interaction. We also find superexchange terms, and an interesting interaction that we characterize as a correlated Kondo effect. There are also additional contributions to these interactions that come from the commutators shown in Eq. \ref{3rd order eq}. We describe these interactions below.

\section{$4^{th}$-Order terms}
\label{sec:4th order}
The expansion was carried out order by order using a Mathematica program written to calculate the many commutators involved, including checks at each order. The first term we find at $4^{th}$-order is the well-known RKKY interaction between the two local sites. In addition, we also find an anti-ferromagnetic superexchange term and a correlated Kondo term. The RKKY interaction can be understood as a $2^{nd}$-order perturbation on the Kondo ground state, with an electron-hole excited state; hence the RKKY term is of $O(J^2/E_F)$. The superexchange process arises from particle-particle and hole-hole excitations, resulting in an excited state that involves either double occupancy or double vacancy on either of the local sites. The electrons hop from both inpurity sites to an intermediate state in the conduction sea, exchange, and hop back. This can only occur when charge fluctuations are allowed at the impurity sites, hence the superexchange term is of $O(J^2/U)$. In the strong Kondo limit ($U \rightarrow \infty$), the superexchange interaction disappears, which is why superexchange is usually not discussed for metallic Kondo systems. 

The correlated Kondo interaction arises from intermediate states that involve an electron of the Kondo singlet hopping from one site to the other and back. Since it also involves doubly-occupied or doubly-vacant excited states on the impurity sites, it is also of $O(J^2/U)$ and will disappear in the large-$U$ limit.

Making the assumption that the constants $J$, $K$ and $I$ are approximately constant over the energy range we are interested in, we make the approximation $J^{\alpha,\alpha '}_{\vec{k},\vec{k'}} \approx J$. $I^{\alpha,\alpha '}_{\vec{k},\vec{k'}}$ is the interaction strength of the double-occupancy term and $K^{\alpha,\alpha '}_{\vec{k},\vec{k'}}$ is the $\epsilon_d$ renormalization energy. Looking at the form of $I$, we can approximate $I^{\alpha,\alpha '}_{\vec{k},\vec{k'}} \approx J/2$, and for the symmetric case where $\epsilon_d \approx U/2$, $K^{\alpha,\alpha '}_{\vec{k},\vec{k'}} \approx J/2$. The expressions for $I^{\alpha,\alpha '}_{\vec{k},\vec{k'}}$ and $K^{\alpha,\alpha '}_{\vec{k},\vec{k'}}$ are also given in App. \ref{app:Higher-S Def}.

The simplified expressions for RKKY and superexchange are then,
\begin{eqnarray}
H_{RKKY} & = & \frac{J^2}{E_f}\frac{(k_F L)^4}{16 \pi} \left[ J_0(k_F R_{12})Y_0(k_F R_{12}) \right. \nonumber \\
 & + & \left. J_1(k_F R_{12})Y_1(k_F R_{12})) \right] \vec{S}_1 \cdot \vec{S}_2
\label{RKKY expr}
\end{eqnarray}
\be
H_{se} = \frac{J^2}{20}\left(\frac{1}{\epsilon} + \frac{1}{U} \right)\frac{(k_F L)4}{(2 \pi)2} \left(\frac{J_1(k_F R_{12})}{k_F R_{12}}\right)^2 \vec{S}_1 \cdot \vec{S}_2
\label{SuEx expr}
\ee
$J_i$ and $Y_i$ are respectively the first and second type Bessel functions of the $i^{th}$-order, and $R_{12}$ is the distance between the two impurities.

\begin{figure}[bht]
\begin{center}
\includegraphics[width=8cm]{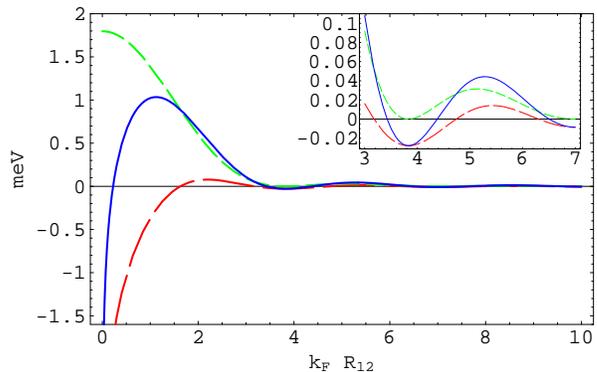}
\end{center}
\caption{Spatial dependence of RKKY and superexchange plotted with $\epsilon_d = -500 \, \mu$eV, $U = 4$ meV, and $E_F = 36$ meV. RKKY is the red (long dashed) curve, superexchange is the green (short dashed) curve, and the blue (solid) curve is the sum of the two. The inset shows the experimentally relevant range of site separation.}
\label{RKKYSuEx_plot}
\end{figure}

The spatial dependence of RKKY and superexchange are plotted in Fig. \ref{RKKYSuEx_plot}, and both show the characteristic $\pi/k_F$ dependence. The ratio of $E_f$ to $U$ was chosen to be 1 in this plot. It must be understood that it is only the sum of RKKY and superexchange that is a physically relevant interaction between the two Kondo impurities. Therefore, by changing the ratio of $E_f$ to $U$, we can favour either the RKKY or superexchange contribution, and hence a more anti-ferromagnetic or more ferromagnetic interaction.

In addition, we also found a correlated Kondo term at $4^{th}$-order, which modulates the Kondo coupling on dot 1 by varying the electron density on dot 2, and vice-versa. The expression for the term is,
\ba
H_{CK1} \!\! & \! = \! & \! J^{1M}J^{2M} \frac{(k_F L)^2}{2 \pi} \left[ \frac{1}{2 |\epsilon_d|} + \frac{1}{3 U} \right] \frac{J_1(k_F R_{12})}{k_F R_{12}} \nonumber \\
 \!\!& \!& \! \sum_{\vec{k},\vec{k}'} (e^{i(\vec{k} + \vec{k'}) \cdot \vec{r}_1} \! + \! e^{-i(\vec{k} + \vec{k'}) \cdot \vec{r}_1}) n_2 \vec{S}_1 \cdot \vec{S}_{m,\vec{k},\vec{k}'} \nonumber \\
H_{CK2} \!\! & \! = \! & \! J^{1M}J^{2M} \frac{(k_F L)^2}{2 \pi} \left[ \frac{1}{2 |\epsilon_d|} + \frac{1}{3 U} \right] \frac{J_1(k_F R_{12})}{k_F R_{12}} \nonumber \\
 \!\!& \!& \! \sum_{\vec{k},\vec{k}'} (e^{i(\vec{k} + \vec{k'}) \cdot \vec{r}_2} \! + \! e^{-i(\vec{k} + \vec{k'}) \cdot \vec{r}_2}) n_1 \vec{S}_2 \cdot \vec{S}_{m,\vec{k},\vec{k}'} 
\ea
The local impurity spins, $\vec{S_1}$ and $\vec{S_2}$, only couple to the conduction electrons in the middle reservoir, $\vec{S}_{m,\vec{k},\vec{k}'} = c_{m,\alpha,\vec{k}}^{\dagger} \frac{\vec{\sigma}_{\alpha \beta}}{2} c_{m,\alpha,\vec{k}'}$, in this case because of the geometry shown in Figure 1. The intermediate excited states of the density-modulated Kondo process involve doubly-occupied and doubly-vacant states on the second dot. Hence, the coupling is of order $O(1/U)$, and depends on the density of the second dot.

\begin{figure}[bht]
\begin{center}
\includegraphics[width=8cm]{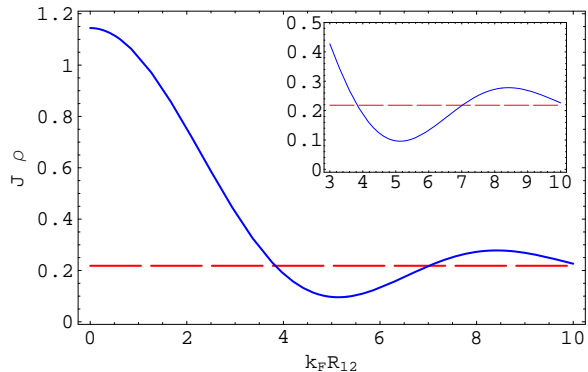}
\end{center}
\caption{Spatial dependence of correlated Kondo term plotted for $\epsilon_d = -500  \, \mu$eV, $U = 4$ meV, and $E_F = 36$ meV, which gives $J \rho = 0.22$. The red (dashed) curve is the Kondo coupling, $J \rho$, and the blue (solid) line is the Kondo coupling plus the correlated Kondo term. The inset, as in Fig. \ref{RKKYSuEx_plot}, shows the experimentally relevant range of site separation. Note that for this interaction, the zero crossing does not occur until nearly $4/k_F$.}
\label{CorKondo_plot}
\end{figure}

The spatial dependence of the correlated Kondo term is plotted in Fig. \ref{CorKondo_plot}, and instead of a $\pi/k_F$ periodicity characteristic of RKKY, it has a $2 \pi/k_F$ periodicity. 

In addition to the above terms, we also found density modulated RKKY and superexchange terms, and also additional scattering terms at $4^{th}$-order. We will ignore the scattering terms as they are of order $O(J/E_F)$ weaker than the scattering terms obtained at $2^{nd}$-order. The complete expressions in terms of the original parameters for RKKY and superexchange are given in App. \ref{app:comp expr}. 

\section{Discussions}
\label{sec:discussions}
The low-energy spin interactions obtained from a generalized Schrieffer-Wolff transformation on an Anderson model of two magnetic impurities include not only the Kondo interaction at $2^{nd}$-order and the well-known RKKY interaction at $4^{th}$-order, but also an antiferromagnetic superexchange interaction, which, unlike the RKKY interaction, cannot be obtained from a perturbation treatment of the Kondo interaction. The superexchange interaction arises from doubly-occupied and hole-hole excitations or double-vacant and particle-particle excitations; hence it cannot be obtained from the Kondo Hamiltonian that does not allow fluctuations in the occupation of the localized states. Therefore, for systems where $U$ is not too large compared to $E_F$, the physically relevant spin interaction between the two impurities is actually the sum of both the RKKY and superexchange interactions. By changing the ratio of $E_f$ to $U$ and the distance between the two impurities, one could potentially tune the system through a phase transition from a Kondo singlet ground state to an impurity singlet ground state. 

Similar processes give rise to the correlated Kondo term that varies the Kondo coupling on dot 1, and hence changes the Kondo temperature, via modulation of the density on dot 2. This allows us to exert a non-local control over the Kondo effect on the first impurity. Due to the short-range nature of this interaction, the two impurities have to be no more than 3 $\lambda_F$ apart for this effect to be observable. This can be achieved in STM experiments on Cu and CuN surfaces where cobalt atoms can be placed as near as 3.6 {\AA} apart. 

These effects can also be experimentally tested in quantum dot systems. In quantum dot systems, the 2 DEG density can be modulated by a back gate voltage, thus changing $\lambda_F$. This would allow us to observe the oscillation of the RKKY and superexchange interaction, and also to tune the exchange coupling between the two impurity spins. The occupation of the local state can also be varied using a separate gate voltage, $V_g$, and the Kondo coupling can be measured using transport measurements through the first dot.

\begin{acknowledgments} 
We would like to thank S. A. Kivelson for enlightening discussions, and for bringing to our attention the use of canonical transformation in the Hubbard model. We would also like to thank D. Goldhaber-Gordon for valuable discussions, and H. D. Chen for assistance in writing the Mathematica program. 

\end{acknowledgments}

\appendix
\section{Kondo Coupling Definitions}
\label{app:Kondo Def}

The definitions for $J^{\alpha,\alpha '}_{\vec{k},\vec{k'}}$, $K^{\alpha,\alpha '}_{\vec{k},\vec{k'}}$ and $I^{\alpha,\alpha '}_{\vec{k},\vec{k'}}$ are given below.
\begin{widetext}
\ba
Ji^{\alpha,\alpha '}_{\vec{k},\vec{k'}} & = & t^{i,\alpha}_{\vec{k}}(t^{i,\alpha '}_{\vec{k'}})^* \left( \frac{1}{\epsilon_{\alpha,\vec{k}} - \epsilon_i} + \frac{1}{\epsilon_{\alpha ',\vec{k'}} - \epsilon_i} - \frac{1}{\epsilon_{\alpha,\vec{k}} - \epsilon_i - U_i} - \frac{1}{\epsilon_{\alpha ',\vec{k'}} - \epsilon_i - U_i} \right) \nonumber \\
Ki^{\alpha,\alpha '}_{\vec{k},\vec{k'}} & = & t^{i,\alpha}_{\vec{k}}(t^{i,\alpha '}_{\vec{k'}})^* \left( \frac{1}{\epsilon_{\alpha,\vec{k}} - \epsilon_i} + \frac{1}{\epsilon_{\alpha ',\vec{k'}} - \epsilon_i} \right) \nonumber \\
Ii^{\alpha,\alpha '}_{\vec{k},\vec{k'}} & = & t^{i,\alpha}_{\vec{k}}(t^{i,\alpha '}_{\vec{k'}})^* \left( \frac{1}{\epsilon_{\alpha,\vec{k}} - \epsilon_i} - \frac{1}{\epsilon_{\alpha,\vec{k}} - \epsilon_i - U_i} \right)
\label{Def cons}
\ea
\end{widetext}

\section{Higher-Order $S$-Operators}
\label{app:Higher-S Def}

The defining equation for $S^{(1)}$ was given in Eq. \ref{S1 Def}, and the defining equations for the higher-order $S^{(i)}$ are,
\begin{widetext}
\ba
(\mathcal{I} - P_G) \ \left( [S^{(2)},H_0] + \frac{1}{2}[S^{(1)},H_I]) \right) P_G & = & 0 \label{2nd order eq} \\
(\mathcal{I} - P_G) \ \left( [S^{(3)},H_0] + \frac{1}{2}[S^{(2)},H_I] + 
\frac{1}{2}[S^{(1)},[S^{(2)},H_0]] + \frac{1}{3}[S^{(1)},[S^{(1)},H_I]] \right) P_G & = & 0 
\label{3rd order eq} \\
(\mathcal{I} - P_G) \ \left( [S^{(4)},H_0] + \frac{1}{2}[S^{(3)},H_I] + 
\frac{1}{2}[S^{(2)}, [S^{(2)},H_0]] + [S^{(2)},\frac{1}{3}[S^{(1)},H_I]] \right. + & & \nonumber \\
\left. \frac{1}{2}[S^{(1)},[S^{(3)},H_0]] + [S^{(1)},\frac{1}{3}[S^{(2)},H_I] +\frac{1}{6}[S^{(1)},[S^{(2)},H_0]] + \frac{1}{8}[S^{(1)},[S^{(1)},H_I]]] \right) \ P_G & = & 0 
\label{4th order eq}
\ea
\end{widetext}
The detailed expressions for the higher-order $S_i$-operators are not shown here as they are too tedious. The effective Hamiltonian up to $4^{th}$-order is then given by the sum of all terms up to and including the $4^{th}$-order expansion.

\section{Complete Expressions}
\label{app:comp expr}

The complete expression for superexchange in terms of the original parameters is given in Eq. \ref{SuEx comp expr}. 
\begin{widetext}
\ba
H_{SuEx} & = & \left[ \frac{1}{12} \sum_{\vec{k},\vec{k1}} \frac{t^{1,m}_{\vec{k}}(t^{1,m}_{\vec{k1}})^*}{\epsilon_{m,\vec{k}} - \epsilon_1} \frac{t^{2,m}_{\vec{k1}}(t^{2,m}_{\vec{k}})^*}{\epsilon_{m,\vec{k1}} - \epsilon_2} \left( \frac{1}{\epsilon_{m,\vec{k}} - \epsilon_1} - \frac{1}{\epsilon_{m,\vec{k}} - \epsilon_1 - U_1} \right) \right. \nonumber \\
 & & + \frac{1}{12} \sum_{\vec{k},\vec{k1}} \frac{t^{1,m}_{\vec{k}}(t^{1,m}_{\vec{k1}})^*}{\epsilon_{m,\vec{k}} - \epsilon_1} \frac{t^{2,m}_{\vec{k1}}(t^{2,m}_{\vec{k}})^*}{\epsilon_{m,\vec{k1}} - \epsilon_2} \left( \frac{1}{\epsilon_{m,\vec{k1}} - \epsilon_2} - \frac{1}{\epsilon_{m,\vec{k1}} - \epsilon_2 - U_2} \right) \nonumber \\
 & & + \frac{1}{24} \sum_{\vec{k},\vec{k1}} \frac{t^{1,m}_{\vec{k}}(t^{1,m}_{\vec{k1}})^*}{\epsilon_{m,\vec{k}} - \epsilon_1} \frac{t^{2,m}_{\vec{k1}}(t^{2,m}_{\vec{k}})^*}{U} \left( \frac{1}{\epsilon_{m,\vec{k}} - \epsilon_1} - \frac{1}{\epsilon_{m,\vec{k}} - \epsilon_1 - U_1} \right) \nonumber \\
 & & + \frac{1}{24} \sum_{\vec{k},\vec{k1}} \frac{t^{1,m}_{\vec{k}}(t^{1,m}_{\vec{k1}})^*}{U} \frac{t^{2,m}_{\vec{k1}}(t^{2,m}_{\vec{k}})^*}{\epsilon_{m,\vec{k1}} - \epsilon_2} \left( \frac{1}{\epsilon_{m,\vec{k1}} - \epsilon_2} - \frac{1}{\epsilon_{m,\vec{k1}} - \epsilon_2 - U_2} \right) \nonumber \\
 & & + \frac{1}{16} \sum_{\vec{k},\vec{k1}} \frac{t^{1,m}_{\vec{k}}(t^{1,m}_{\vec{k1}})^* t^{2,m}_{\vec{k1}}(t^{2,m}_{\vec{k}})^*}{\epsilon_{m,\vec{k1}} - \epsilon_2 - U_2} (\frac{1}{\epsilon_{m,\vec{k}} - \epsilon_1} + \frac{1}{\epsilon_{m,\vec{k1}} - \epsilon_2}) \left( \frac{1}{\epsilon_{m,\vec{k}} - \epsilon_1} - \frac{1}{\epsilon_{m,\vec{k}} - \epsilon_1 - U_1} \right) \nonumber \\
 & & + \left. \frac{1}{16} \sum_{\vec{k},\vec{k1}} \frac{t^{1,m}_{\vec{k}}(t^{1,m}_{\vec{k1}})^* t^{2,m}_{\vec{k1}}(t^{2,m}_{\vec{k}})^*}{\epsilon_{m,\vec{k}} - \epsilon_1 - U_1} (\frac{1}{\epsilon_{m,\vec{k}} - \epsilon_1} + \frac{1}{\epsilon_{m,\vec{k1}} - \epsilon_2}) \left( \frac{1}{\epsilon_{m,\vec{k1}} - \epsilon_2} - \frac{1}{\epsilon_{m,\vec{k1}} - \epsilon_2 - U_2} \right) \right] \nonumber \\
 & & \times \left( e^{i(\vec{k} - \vec{k1}) \cdot \vec{R_{12}}} +  e^{-i(\vec{k} - \vec{k1}) \cdot \vec{R_{12}}} \right) \vec{S}_1 \cdot \vec{S}_2
\label{SuEx comp expr}
\ea
\end{widetext}

The complete expression for RKKY is shown in Eq. \ref{RKKY comp expr}.

\begin{widetext}
\ba
H_{RKKY} & = &  \frac{1}{8} \left[\sum_{\vec{k},\vec{k1}} \frac{t^{1,m}_{\vec{k}}(t^{1,m}_{\vec{k1}})^*}{\epsilon_{m,\vec{k}} - \epsilon_1} \frac{t^{2,m}_{\vec{k1}}(t^{2,m}_{\vec{k}})^*}{\epsilon_{m,\vec{k1}} - \epsilon_1} \left( \frac{1}{\epsilon_{m,\vec{k}} - \epsilon_2} + \frac{1}{\epsilon_{m,\vec{k1}} - \epsilon_2} - \frac{1}{\epsilon_{m,\vec{k}} - \epsilon_2 - U_2} - \frac{1}{\epsilon_{m,\vec{k1}} - \epsilon_2 - U_2} \right) \right. \nonumber \\
 & & \left. + \sum_{\vec{k},\vec{k1}} \frac{t^{1,m}_{\vec{k}}(t^{1,m}_{\vec{k1}})^*}{\epsilon_{m,\vec{k}} - \epsilon_2} \frac{t^{2,m}_{\vec{k1}}(t^{2,m}_{\vec{k}})^*}{\epsilon_{m,\vec{k1}} - \epsilon_2} \left( \frac{1}{\epsilon_{m,\vec{k}} - \epsilon_1} + \frac{1}{\epsilon_{m,\vec{k1}} - \epsilon_1} - \frac{1}{\epsilon_{m,\vec{k}} - \epsilon_1 - U_1} - \frac{1}{\epsilon_{m,\vec{k1}} - \epsilon_1 - U_1} \right) \right] \nonumber \\
 & & \times \left( e^{-i(\vec{k} - \vec{k1}) \cdot \vec{R_{12}}} + e^{i(\vec{k} - \vec{k1}) \cdot \vec{R_{12}}} \right) \vec{S}_1 \cdot \vec{S}_2
\label{RKKY comp expr}
\ea
\end{widetext}

The complete expression for the correlated Kondo term is too involved and will not be reproduced here, but is available from the authors upon request.

\bibliographystyle{btxbst}
\bibliography{Kondo}

\begin{thebibliography}{10}

\bibitem{JKondo}
J.~Kondo,
\newblock Prog. Theoretical Physics {\bf 32}, 37 (1964).

\bibitem{Hewson}
A.~C. Hewson,
\newblock {\em The Kondo Problem to Heavy Fermions, Cambridge Studies In
  Magnetism Vol. 2},
\newblock Cambridge University Press, Cambridge, England, 1993.

\bibitem{DGG98Sc}
D.~G.-G. et~al.,
\newblock Science {\bf 391}, 156 (1998).

\bibitem{DGG98PRl}
D.~G.-G. sl~et al.,
\newblock Phys. Rev. Lett. {\bf 81}, 5225 (1998).

\bibitem{Crommie}
V.~Madhavan, W.~Chen, T.~Jamneala, M.~Crommie, and N.~S. Wingreen,
\newblock Science {\bf 280}, 567 (1998).

\bibitem{Jones87}
B.~A. Jones and C.~M. Varma,
\newblock Phys. Rev. Lett. {\bf 58}, 843 (1987).

\bibitem{Affleck}
I.~Affleck, A.~W.~W. Ludwig, and B.~A. Jones,
\newblock Phys. Rev. B {\bf 52}, 9528 (1995).

\bibitem{Gan}
J.~Gan,
\newblock Phys. Rev. B {\bf 51}, 8287 (1995).

\bibitem{Marcus}
N.~J. Craig, J.~M. Taylor, E.~A. Lester, M.~P.~H. C.~M.~Marcus, and A.~C.
  Gossard,
\newblock Science {\bf 304}, 565 (2004).

\bibitem{PALee88}
T.~K. Ng and P.~A. Lee,
\newblock Phys. Rev. Lett. {\bf 61}, 1768 (1988).

\bibitem{PALee93}
Y.~Meir, N.~S. Wingreen, and P.~A. Lee,
\newblock Phys. Rev. Lett. {\bf 70}, 2601 (1993).

\bibitem{Wingreen94}
J.~R. Schrieffer and P.~Wolff,
\newblock Phys. ~Rev {\bf 149}, 491 (1966).

\bibitem{Schrieffer-Wolff}
J.~R. Schrieffer and P.~Wolff,
\newblock Phys. ~Rev {\bf 149}, 491 (1966).

\bibitem{MacDonald}
A.~H. MacDonald, S.~M. Girvin, and D.~Yoshioka,
\newblock Phys. Rev. B {\bf 37}, 9753 (1988).

\end{thebibliography}

\end{document}